\documentclass[a4paper,twocolumn,11pt,accepted=2017-05-09]{quantumarticle}
\pdfoutput=1
\usepackage[utf8]{inputenc}
\usepackage[english]{babel}
\usepackage[T1]{fontenc}
\usepackage{amsmath}
\usepackage{hyperref}
\usepackage{tikz}
\usepackage{lipsum}
\usepackage{amsmath,amssymb,graphicx}
\usepackage{ulem}
\usepackage{subfigure}
\usepackage{color}
\usepackage{float}
\usepackage{amsfonts}
\usepackage{bm}
\usepackage{mathtools}

\newcommand{\f}{\varphi}

\renewcommand{\Im}{{\rm Im}}
\newcommand{\ii}{{\rm i}}
\newcommand{\dd}{{\rm d}}

\begin{document}

% \title{The QFOCUS Project}
% \title{Optimal Quantum Focusing with Cylindrical Shells}
% \title{Quantum Particle-Collecting Nanotubes}
% \title{Quantum Particles Sink}
\title{Nanotubes as Sinks for Quantum Particles}
% \title{Matter-Wave Trapping with Cylindrical Shells}

\author{Constantinos Valagiannopoulos}
\affiliation{Department of Physics, Nazarbayev University, Nur-Sultan 010000, Kazakhstan}
\orcid{0000-0003-1560-2576}
\thanks{\texttt{valagiannopoulos@gmail.com}.}
\maketitle

\begin{abstract}
Nanotubes with proper thickness, size and texture make ultra-efficient sinks for the quantum particles traveling into specific background media. Several optimal semiconducting cylindrical layers are reported to achieve enhancement in the trapping of matter waves by 2-3 orders of magnitude. The identified shells can be used as pieces in quantum devices that involve the focusing of incident beams from charge pumps and superconducting capacitors to radiation pattern controllers and matter-wave lenses.    
\end{abstract}

\section{Introduction}
Concentrating the electromagnetic fields intensity across a region of space is a generic requirement in several photonic operations like the boost of nonlinearities for generating higher harmonics \cite{HighHarmonicGeneration} or the enhancement of fluorescence molecular radiation via coupling to inorganic nanoparticles \cite{EnhancementQuenching}. Surface plasmon modes, developed along metal/dielectric interfaces, are also routinely used as vessels for the localization of energy for sensing and waveguiding purposes \cite{PlasmonicsLocalization}. In addition, lenses make a significant category of devices requiring signal focusing to realize subwavelength resolution imaging and dynamic holography with help from patterned surfaces \cite{MetalensesVisible} or controllable active elements \cite{ReconfigurableActive}. Furthermore, absorption is another form of power collection into specific volumes that is accomplished by metamaterial resonators reaching near-unity performance \cite{PerfectMetamaterial} or conjugately-matched layers breaking the black-body limits \cite{MyJ66}.

The trapping of incoming particles can be also seen as intensity concentration; indeed, the manipulation of nanostructures via cooling of atoms or  momentum transfer has substantially affected nanotechnology \cite{OpticalManip} and biophysics at the nanoscale \cite{OptTrap}. Perfect light trapping has been also achieved by artificially increasing the path length of the incident rays \cite{MyJ111} while an optical black hole has been demonstrated by bending the beam trajectories with use of anisotropic and inhomogeneous materials \cite{EnhancedControl}. Similar investigations are provided for cylindrical \cite{CylindricalBlackHole} and spherical \cite{FDTDAnalysis} mantles simulated to effectively absorb the incident waves from all directions or gradient index claddings fabricated to operate at microns wavelengths \cite{BroadbandOMNI}.

Quantum analogues of these optical traps have been examined to report formation of matter waves by magnetically tuning the interactions in Bose-Einstein condensates \cite{FormationProp} and nonlinear self-trapping of particles in periodic potentials \cite{NonlinearSelfTrapping}. Interestingly, the stability of similar trapping effects has been thoroughly studies \cite{StabilityStanding}, while coherent perfect absorption of quantum signal has been experimentally demonstrated by identified the dissipative regime of the setup \cite{CoherentPerfect}. Moreover, heat machines response is found substantially assisted by quantum absorption \cite{QuantumAbsorption} while electron trapping is illustrated in semi-conducting nanowires \cite{ElectronTrapping} and tunnel barriers are reported to act like sinks for impinging quantum particles \cite{MacroscopicQuantum}. 

Nanotubes, namely hollow cylinders at the nanoscale, constitute a unique class of materials extensively employed in photonic and quantum designs that serve a broad range of utilities from solar energy collection and electromagnetic shielding to matter wave switching and quantum signal processing \cite{ScalableSynthesis, 1DNanotubes}. They are usually operated collectively in large clusters with almost parallel axes making them suitable to work as carbon supercapacitors and actuators \cite{CarbonNanotubes} due to their beneficial electromechanical properties or towards rapid accumulation of generated carriers securing high power conversion rates \cite{NanowireDyeSensitized}. As far as their fabrication are concerned, nanotubes can be synthesized by plasma-enhanced chemical vapor deposition if patterned growth is required \cite{3DMacroporous} or via electric charge discharge where vaporization of the cladding medium leads to cylindrical formations \cite{FabReview}. Importantly, carbon nanotubes and other nano-coatings are also commercially available demonstrating their potential to be used in integrating systems with several applications \cite{NanoWorld}.

In this work, we optimize the dimensions of nanotubes to work as sinks for impinging quantum particles. Several combinations of media, picked from a long list with their effective macroscopic properties, are tested and the probability to find the particle at the core of the considered cylindrical structure enhances by 2-3 orders of magnitude. The reported effect is very sensitive to changes in the energy of matter waves admitting the optimal setups to function as sharp filters or sensors \cite{QEInvited, MyJ101}. On the contrary, the response is very robust with respect to fabrication defects involving the material of the cladding, while the nature of the sustained resonances is unveiled by inspection of the spatial distribution of the wave function and the probability current. All these high-performing matter-wave sinks may offer additional degrees of freedom in the design of quantum devices calling for focusing and concentration of particle beams such as charge pumps, parametric amplifiers or superconducting capacitors.

\begin{figure}[ht!]
\centering
{\includegraphics[width=6.0cm]{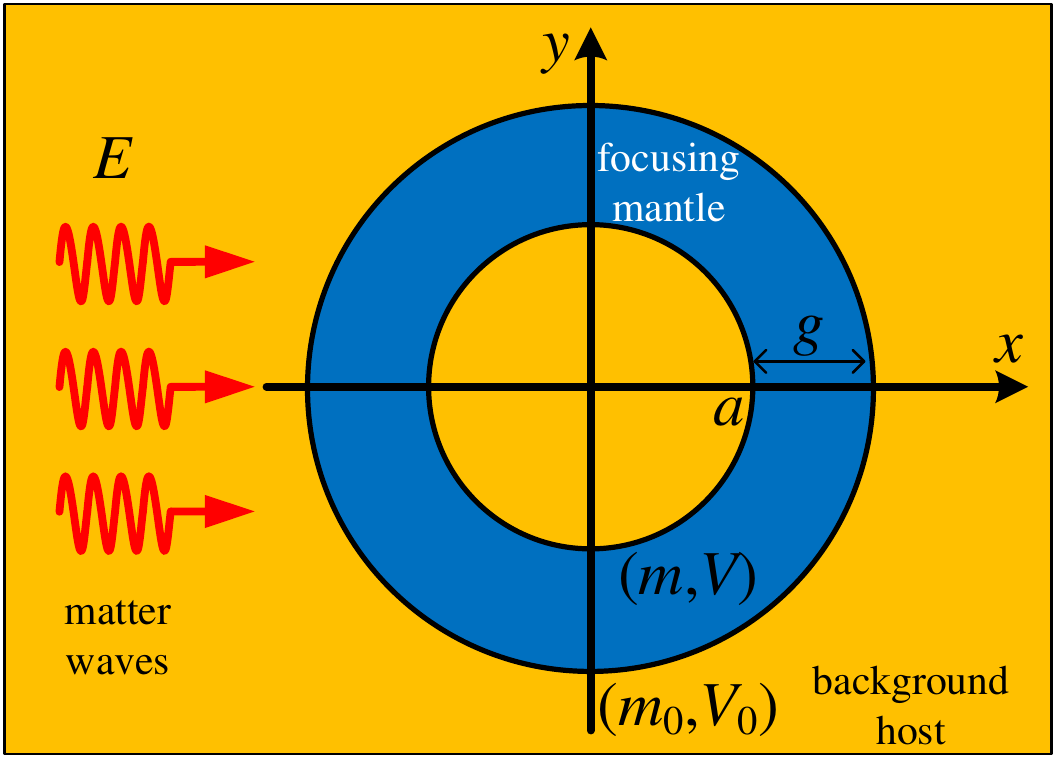}}
\caption{The physical configuration of the cylindrical mantle in the presence of incoming matter plane waves. The quantum signal is concentrated across the internal core since it is collected and focused by the cladding.}
\label{fig:Fig1}
\end{figure}

\section{Mathematical Formulation}

\subsection{Configuration and Objective}
We consider a cylindrical mantle of internal radius $a$ and thickness $g$ illuminated by plane matter waves of energy $E$. The background medium is characterized by effective mass $m_0$ and a local potential energy level $V_0$, while the electrons inside the shell behave like particles of mass $m$ and are subjected to potential field of magnitude $V$. All these quantities are indicated in Fig. \ref{fig:Fig1} and the Cartesian coordinate system $(x,y,z)$ is shown which corresponds to the equivalent cylindrical coordinate system $(r,\f,z)$. One can define the wavelength of the incoming particle beam as: $\lambda_0\equiv 2\pi/k_0=2\pi\hbar/\sqrt{2 m_0 E}$ and the wavelength into the mantle as: $\lambda\equiv 2\pi/k=2\pi\hbar/\sqrt{2 m (E+\Delta V)}$, where $\Delta V \equiv V_0-V$ and $\hbar$ is the reduced Plank constant.

Our objective would be to optimally select the sizes $(a,g)$ for a given pair of materials expressed through the triplet of parameters $\{m_0,m,\Delta V\}$, so that the incoming matter wave is maximally focused across the internal disk $r<a$. 

\subsection{Rigorous Solution}
It is well-known that a matter plane wave, representing equal probability of particle existence across planes normal to the propagation direction $+x$, is given as follows:
\begin{eqnarray}
\Psi_{inc}(r,\f)=\sum_{n=-\infty}^{+\infty}(-1)^n J_n(k_0 r)e^{\ii n \f},
\label{PsiInc}
\end{eqnarray}
where $J_n$ is the Bessel function of order $n$. If one expands the solution of Schrödinger equation $\nabla\cdot\left[\frac{1}{m(\textbf{r})}\nabla\right]\Psi(\textbf{r})+\frac{E-V(\textbf{r})}{\hbar^2/2}\Psi(\textbf{r})=0$ in the corresponding eigenfunctions of each region and demand for continuity of $\Psi$ and $\frac{1}{m}\partial \Psi/\partial{r}$ across the separating boundaries $r=a,(a+g)$, the wave function inside the core ($r<a$) is rigorously found as:
\begin{eqnarray}
\Psi(r,\f)=\sum_{n=-\infty}^{+\infty}(-1)^n C_n J_n(k_0 r)e^{\ii n \f},
\label{PsiInt}
\end{eqnarray}
where $C_n$ are well-determined complex coefficients and $\textbf{r}$ is the position vector.

\subsection{Focusing Metric}
Our aim is to identify the conditions under which maximal concentration of quantum signal is achieved into the core of the regarded configuration. Thus, a good metric to evaluate the strength of this effect would be the integral of probability $|\Psi|^2$ from \eqref{PsiInt} across the disk ($r<a$) over the same quantity in the absence of cladding, namely, $\int_0^{2\pi}\!\!\! \int_0^a |\Psi_{inc}|^2 r \dd r \dd \f=\pi a^2$. The ratio can be defined as follows:
\begin{eqnarray}
\!\!\!\! Q\!\equiv\!\frac{\int_0^{2\pi}\!\!\! \int_0^a |\Psi|^2       r \dd r \dd \f}
              {\int_0^{2\pi}\!\!\! \int_0^a |\Psi_{inc}|^2 r \dd r \dd \f}
						\!=\!\sum_{n=-\infty}^{+\infty}|C_n|^2F_n,
\label{FocMetric}
\end{eqnarray}
where:
\begin{eqnarray}
F_n=\nonumber\\
J^2_{n-1}(k_0a)\!+\!J^2_n(k_0a)\!-\!2n\frac{J_{n-1}(k_0a)J_n(k_0a)}{k_0a}.
\label{FFunction}
\end{eqnarray}

Once the size of the device is small compared to the incident wavelength $\lambda_0$, namely, if $k_0a,k_0g\ll 1$, only the central term from \eqref{FocMetric} is significant and thus the metric $Q$  behave like:
\begin{eqnarray}
Q \cong |C_0|^2 F_0 \sim \frac{k_0a}{J^2_0(k_0a)}.
\label{ApproxQ}
\end{eqnarray}

\begin{figure}[ht!]
\centering
\subfigure[]{\includegraphics[width=5.5cm]{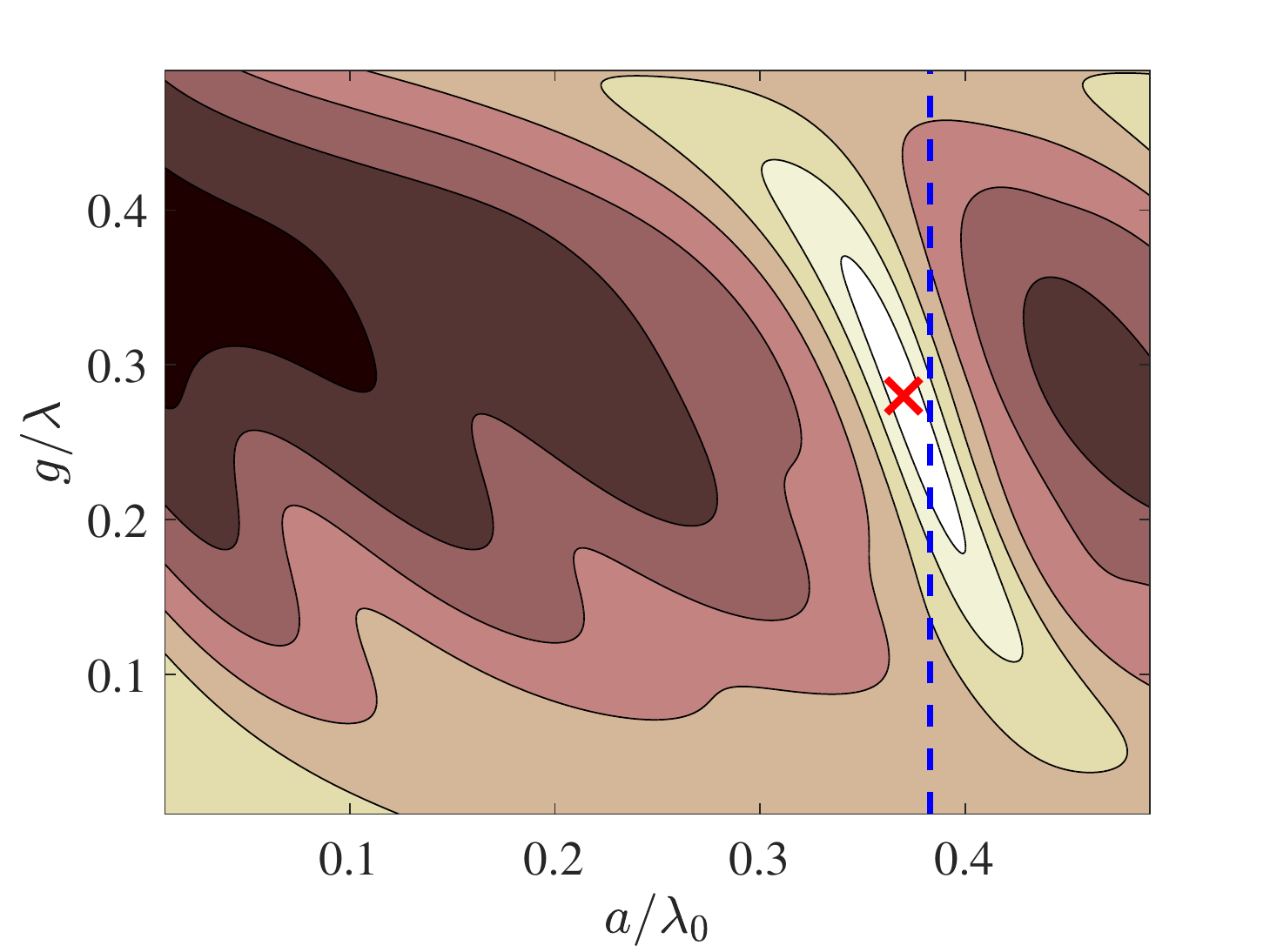}
\label{fig:Fig12}}
\subfigure[]{\includegraphics[width=5.5cm]{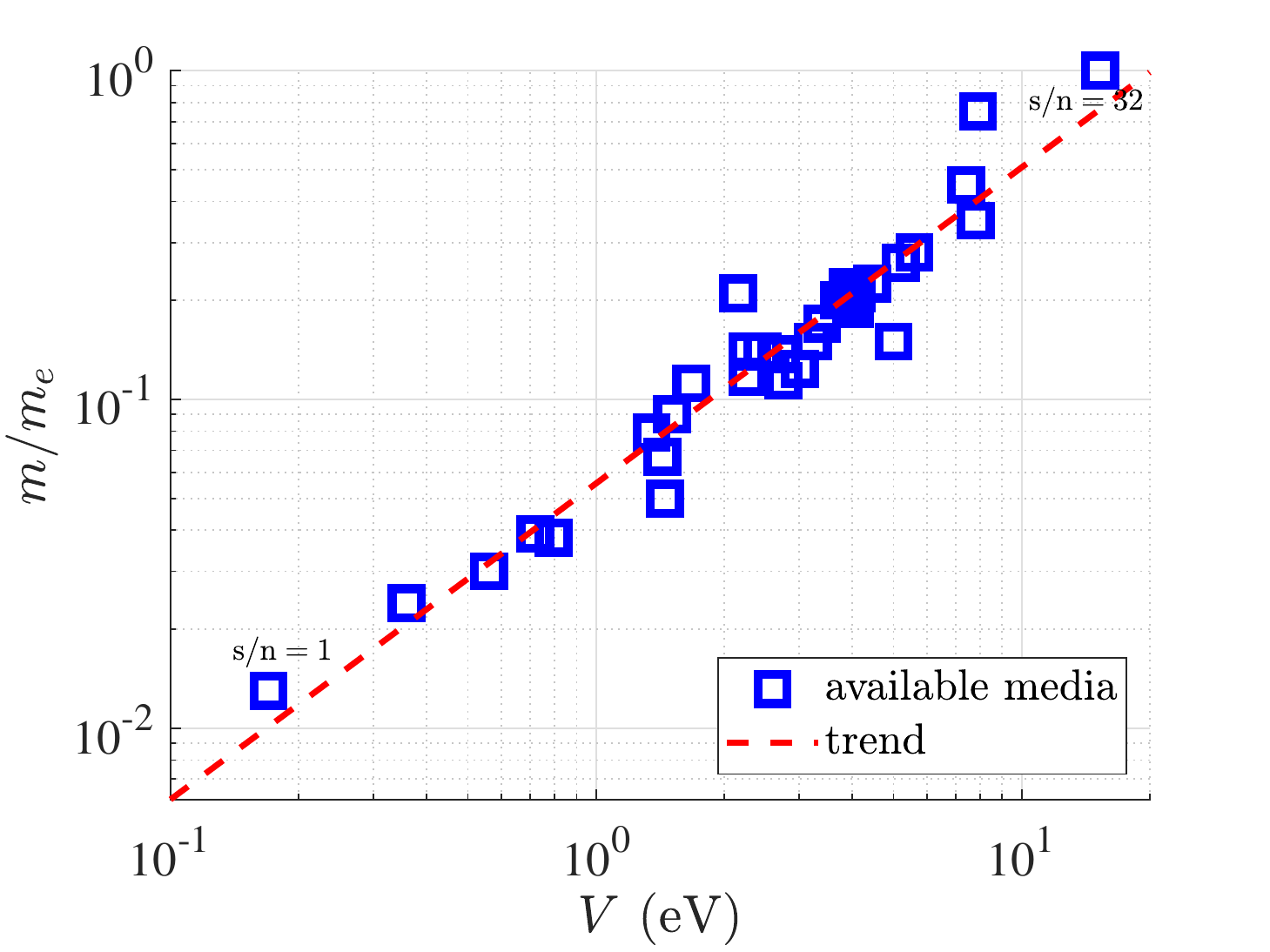}
\label{fig:Fig12b}}
\caption{(a) Typical variation of focusing metric $Q$ from \eqref{FocMetric},\eqref{FFunction} across internal radius $0<a<\lambda_0/2$ and mantle thickness $0<g<\lambda/2$. The blue dashed line gives the value $a=z_0\lambda/(2\pi)$ as indicated by \eqref{ApproxQ} and the red $\times$ marker corresponds to the optimal configuration. Boundary maxima are automatically rejected. InSb cladding in InN background. (b) Correlation between potential energy $V$ and effective mass for the available $U=32$ media each of which is assigned a serial number (s/n) after sorted in ascending order of their potential level. $m_e$ is the inertial mass of the particles.}
\end{figure}

\section{Numerical Simulations}

\subsection{Optimization Strategy}
We have compiled a long list of available media containing their effective masses and potential energy levels; each of the records has been assigned a serial number (s/n) indicating how high is its potential in ascending order. Our strategy will be to try every single of them as a candidate of mantle material $(m,V)$ that creates the focusing at a specific background texture $(m_0,V_0)$, chosen from the same list. We assume $\Delta V>0\Rightarrow V_0>V$ to secure the propagating nature of matter waves into the shell, regardless of the impinging energy $E$. 

For each couple of quantum media, we will run a secondary optimization on the radii map scanning all possible values kept below the local half wavelengths. Indeed, since the considered media are lossless ($m_0,m>0$) and evanescent behavior is ruled out ($\Delta V>0$), the dynamic effects ($\Psi$) are periodic with periods $\lambda_0$ and $\lambda$ in the background and the mantle, respectively. Therefore, we maximize the focusing metric $Q$ from \eqref{FocMetric},\eqref{FFunction} across the rectangular domain defined by $0<a<\lambda_0/2$ and $0<g<\lambda/2$ (since we care about $|\Psi|^2$. Importantly, we avoid to pick boundary extrema since our intention is to encapsulate maxima ``in-the-box'' instead of reporting large $Q$ values of resonances occurring outside of the regarded parametric $(a,g)$ map \cite{MyJ90}. The optimal radius $a$ of the internal disk will be found close to $a\cong z_0/k_0$ where $z_0>0$ is the smallest positive zero of equation $J_0(z)=0$, as imposed by \eqref{ApproxQ}. 

A typical variation of the metric $Q$ across the map of dimensions $(a,g)$ is depicted in Fig. \ref{fig:Fig12}, where the blue vertical dashed line denotes the radius $a=z_0/k_0$ and the red $\times$ marker corresponds to the detected maximum. It is clear that the most substantial magnitudes emerge close to this dashed line while the optimal operation point is found away from $g=0,\lambda/2$ which indicates a quasi-periodic behavior with respect to $g$ of period $\lambda/2$. Similarly, the maxima re-appear for larger internal radii $a$ with period $\lambda_0/2$; in this sense, our optimization captures the primary resonance corresponding to the physically smaller quantum sink. We pay particular attention in reporting maxima that are not boundary extrema, namely, they solely belong to the domains defined by the imposed radii ranges \cite{MyJ93}. 

In Fig. \ref{fig:Fig12b}, we represent the various media of our list as combinations of potential levels $V$ in eV, expressing the minimal energy needed to extract an electron from the medium into vacuum, and effective masses $m$ in terms of inertial particle mass $m_e$, expressing the slope of the bands for Floquet-Bloch waves created into periodic crystalline texture. Together with the discrete points, each of which corresponds to one material of specific serial number (s/n) indicating how high is the local potential, we show the correlation trend which is almost linear (dashed line) and  upward sloping. Numerous ($U=32$) alternative materials are depicted and it is clear that, with the regarded set of media, a significant part of the parametric space of the quantum features $(V,m/m_e)$ is covered coherently. Thus, the optimization with respect to the respective textural pairs is natural to give results close to the global optima.

\begin{figure}[ht!]
\centering
\subfigure[]{\includegraphics[width=5.5cm]{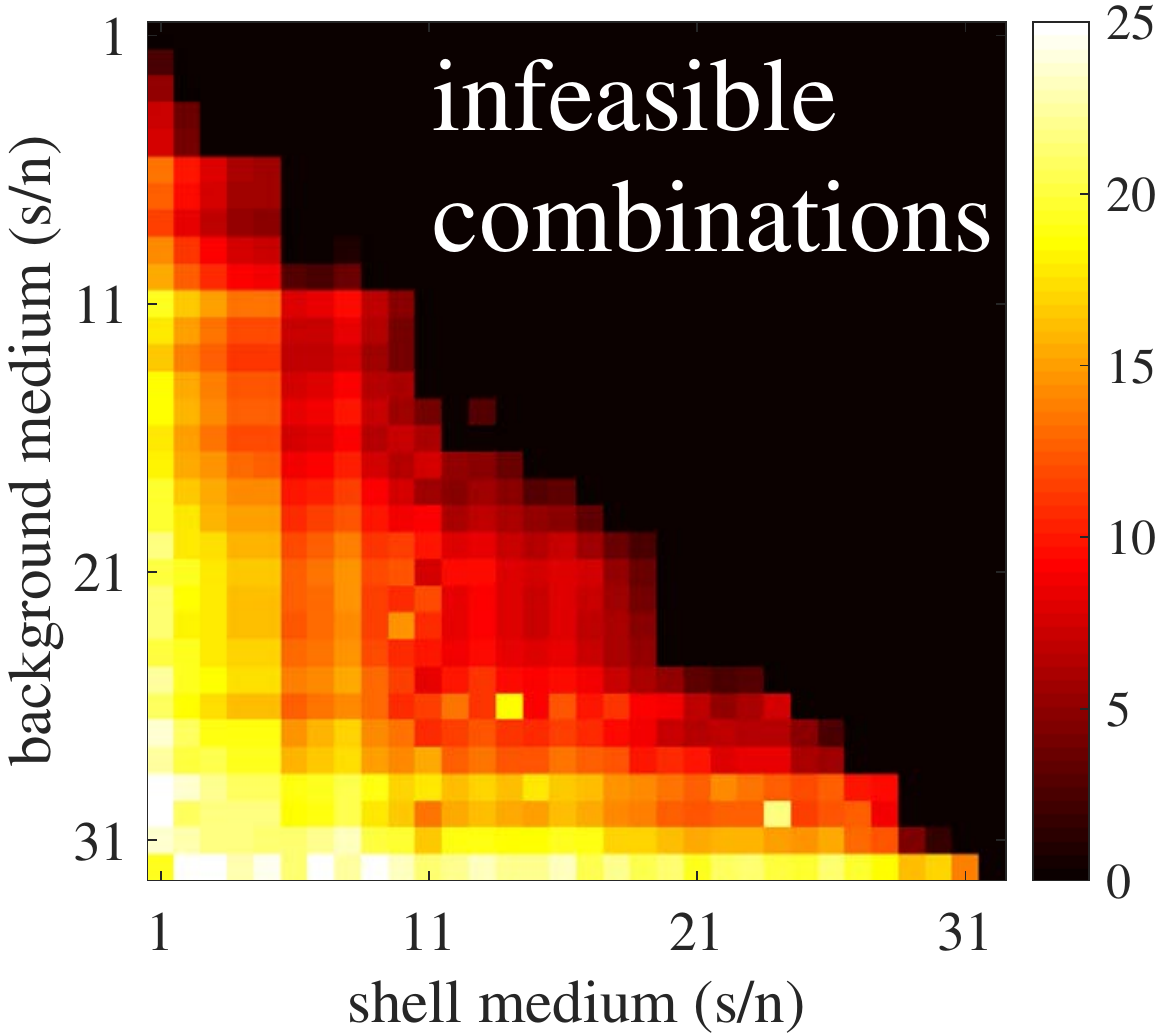} 
   \label{fig:Fig2a}}
\subfigure[]{\includegraphics[width=5.5cm]{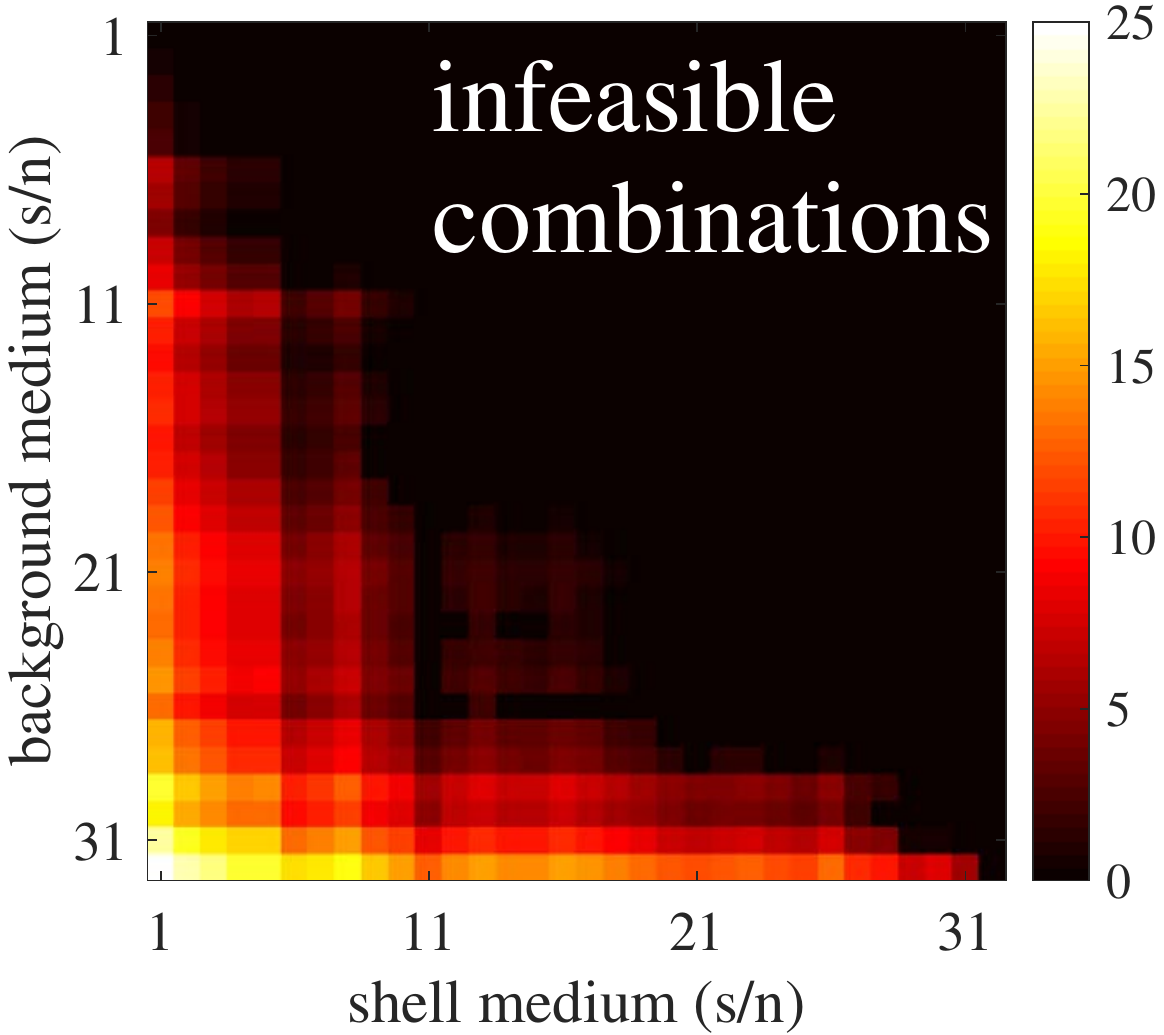}
   \label{fig:Fig2b}}
\caption{Maximal focusing metric $Q$ (in dB) from \eqref{FocMetric},\eqref{FFunction} for each combination of background media (rows) and shell materials (columns). The materials are sorted with ascending order of potential level via their serial number (s/n) for two different impinging energies: (a) $E=0.1$ eV, (b) $E=1$ eV. Only cases with $\Delta V=V_0-V>0$ are considered, otherwise the combination of materials is infeasible.}
\label{fig:Figs2}
\end{figure}

In Fig. \ref{fig:Figs2}, we show the maxima $Q$ in dB obtained via the optimization process described above; along the vertical axis we indicate the serial number of the background material and along the horizontal axis we represent the s/n of the cladding medium. In this way, we formulate a $U\times U$ rectangular matrix of $U=32$ different media presented in Fig. \ref{fig:Fig12b}, that give more than a thousand alternative setups. For each of the shown pixels, an optimization like that of Fig. \ref{fig:Fig12} is performed; obviously, the upper right triangle corresponds to $\Delta V<0$ and, thus, we assign the label ``infeasible combinations'' according to the aforementioned specifications. 

In Fig. \ref{fig:Fig2a}, we regard low-energy particles ($E=0.1$ eV) and the enhancement can be as large as 25 dB if the potential contrast between the background and the cladding is substantial. Note that once the energy level of the host material $V_0$ decreases and that of the shell medium $V$ increases, the maximal metric $Q$ gets smaller but usually remains much higher than 10 dB. In this way, several optimal configurations are proposed each of which may make an ultra-efficient nanotube sink in various quantum setups. In Fig. \ref{fig:Fig2b}, we investigate electrons of higher energy ($E=1$ eV) and, despite the fact that the focusing performance is still very high, it gets diminished compared to the case of Fig. \ref{fig:Fig2a}. The best score is recorded at the lower left pixel where the background has the maximum potential (s/n=$U$=32, diamond) and the shell the minimum one (s/n=1, indium antimonide), from the palette of available materials.

\begin{figure}[ht!]
\centering
\subfigure[]{\includegraphics[width=5.5cm]{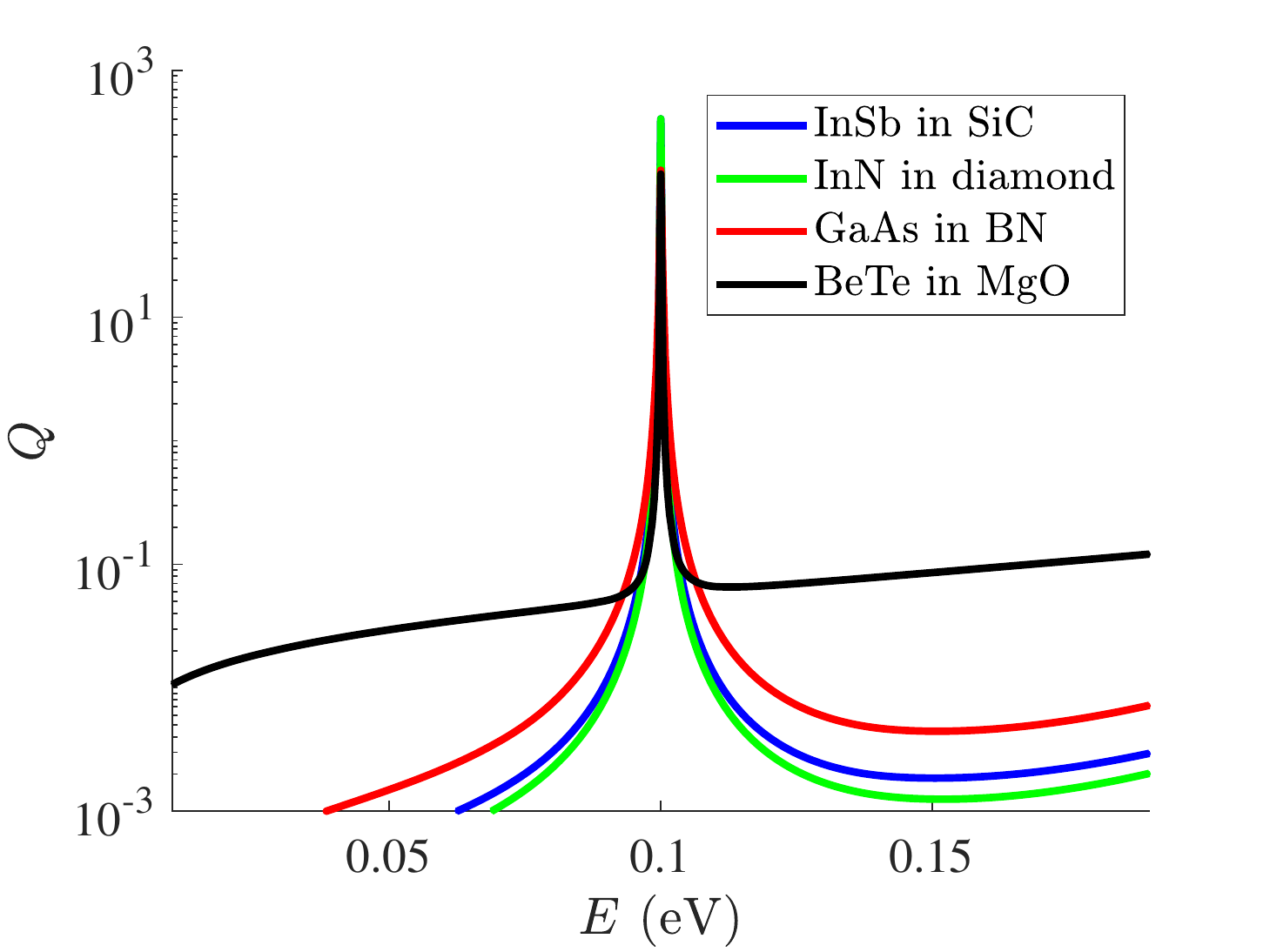}
   \label{fig:Fig3a}}
\subfigure[]{\includegraphics[width=5.5cm]{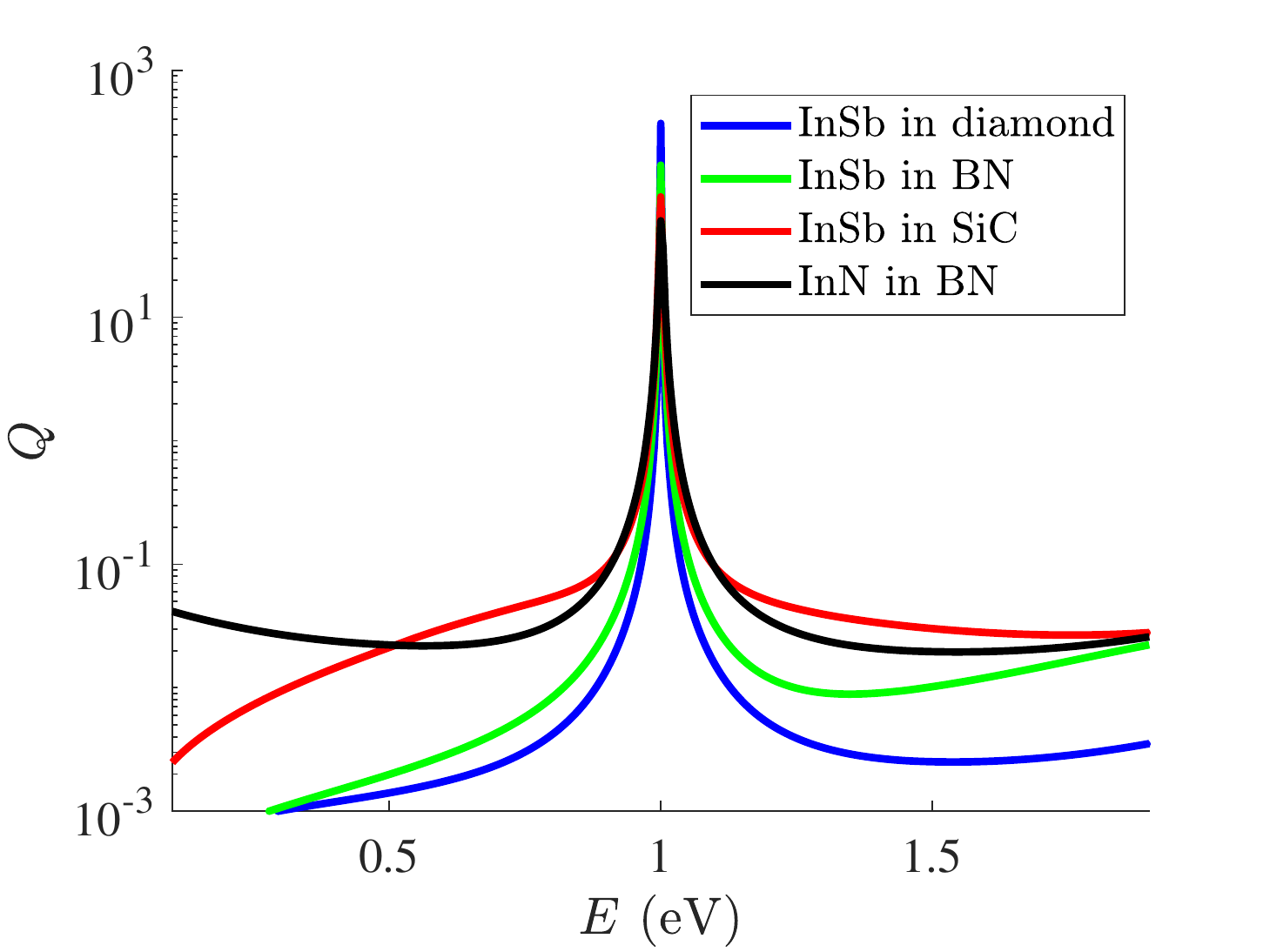}
   \label{fig:Fig3b}}
\caption{Focusing metric $Q$ from \eqref{FocMetric},\eqref{FFunction} as a function of energy $E$ of impinging particles for several optimized designs from Fig. \ref{fig:Figs2} with: (a) central energy $E=0.1$ eV, (b) central energy $E=1$ eV.}
\label{fig:Figs3}
\end{figure}

\subsection{Maximally Trapping Setups}
It would be meaningful to understand the dependence of the trapping effect on the energy of impinging particles; thus, we choose some of the most highly-focusing nanotubes and represent the metric $Q$ as a function of $E$. In Fig. \ref{fig:Fig3a}, we pick four of the best sinks from Fig. \ref{fig:Fig2a} (operated for low-energy particles) and we observe a remarkable energy selectivity for all of the designs, resembling the ultra-sharp responses with respect to the incidence angle in quantum Fabry-Perot resonators \cite{QFP}. It is remarkable that with a change by one percent in the particles energy, the performance $Q$ drops by two orders of magnitude; therefore, the proposed layouts can serve filtering, sensing and switching objectives in a variety of quantum devices. In Fig. \ref{fig:Fig3b}, we consider high-energy particles and the responses from four of the most successful setups of Fig. \ref{fig:Fig2b} are again depicted as functions of energy $E$. Once more time, extremely high sensitivity is exhibited while the drop of $Q$ from the central operating energy is more abrupt for the more highly trapping nanotubes, as in Fig. \ref{fig:Fig3a}.

When it comes to the materials employed, those with very high potential energy $V_0$, like diamond, boron nitride or silicon carbide are befitted for background hosts while media of $V\ll V_0$, like indium-based semiconductors, are more suitable as claddings. These observations are compatible with the texture of state-of-the-art quantum designs routinely employing similar substances to build repeaters comprising optically active spin qubits \cite{QMAT1}, luminescent emitters made of quantum dots \cite{QMAT2}, narrowband quantum light sources \cite{QMAT3} and nanoscale sensors \cite{QMAT4}.       
 
\begin{figure}[ht!]
\centering
\subfigure[]{\includegraphics[width=5.5cm]{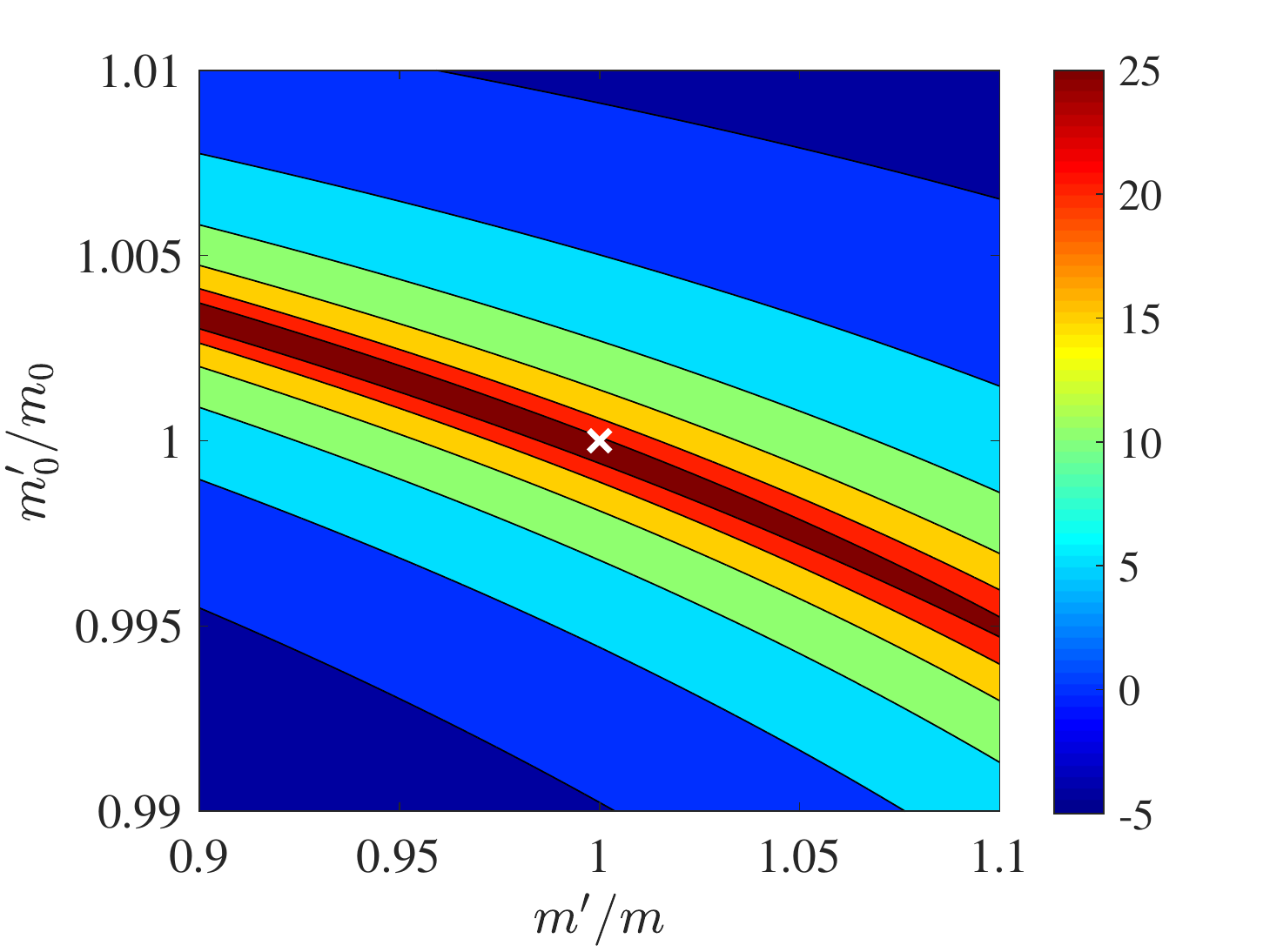}
   \label{fig:Fig4a}}
\subfigure[]{\includegraphics[width=5.5cm]{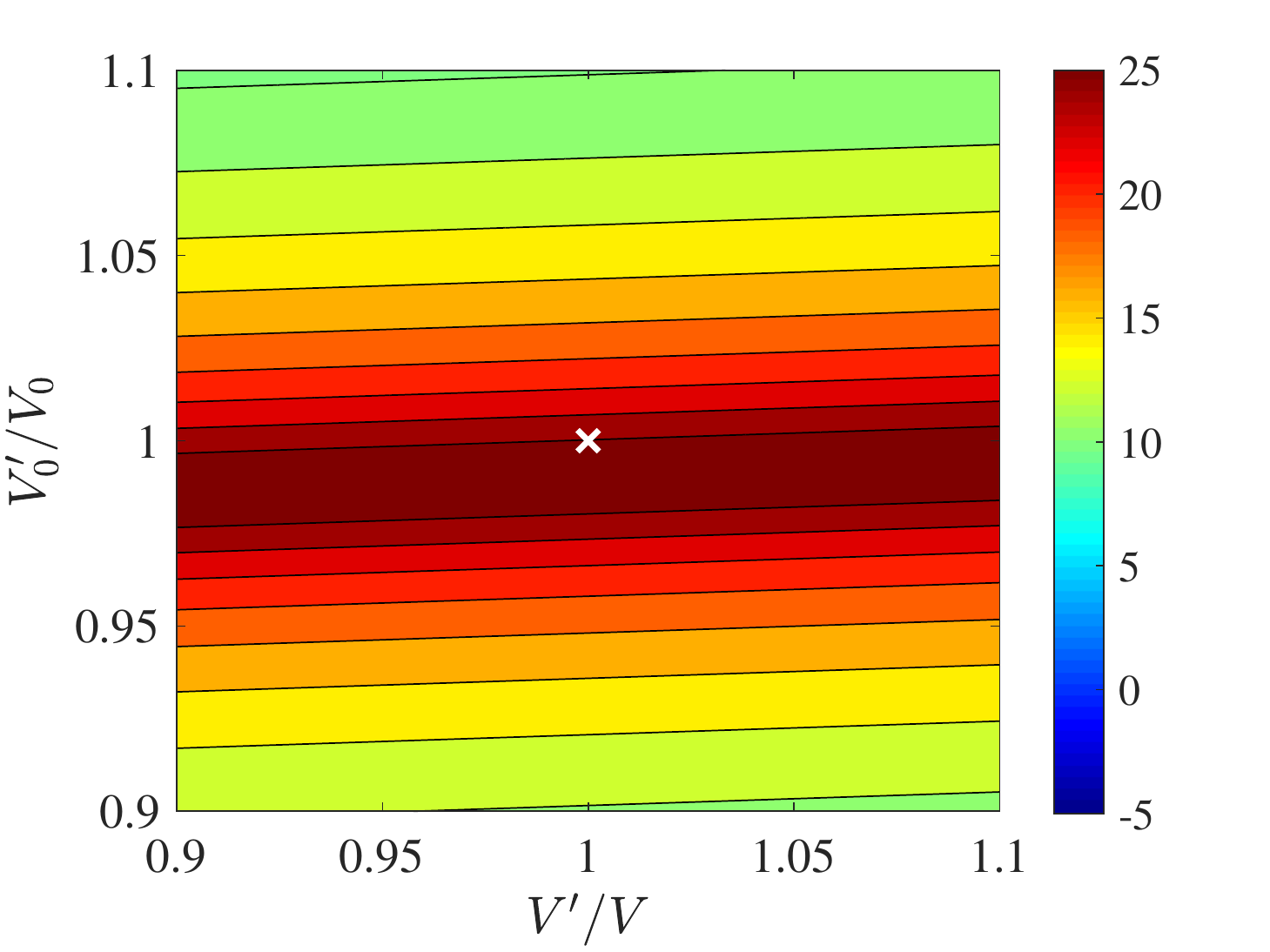}
   \label{fig:Fig4b}}
\caption{Focusing metric $Q$ for: (a) misselected estimation of effective masses and (b) misengineered potential levels into the background medium (vertical axis) and into the cladding (horizontal axis). InN shell into diamond background, low-energy particles: $E=0.1$ eV.}
\label{fig:Figs4}
\end{figure}

\subsection{Performance Robustness}
Quantum interactions are very vulnerable to noise which constitutes the major obstacle in performing reliable quantum signal processes since not only the magnitude but also the phase of the wave functions is affected \cite{Qnoise}. Similarly, given the nanometer-sized wavelengths of the developed matter waves, designs with extremely thin layers are required and, accordingly, the estimation of macroscopic parameters describing each medium is critical. Thus, it is meaningful to examine the response of the device when the effective masses $(m_0,m)$ and the potential levels $(V_0,V)$ are not well-calculated. 

In Fig. \ref{fig:Fig4a}, we show the trapping metric $Q$ for misselected estimation $(m'_0,m')$ of effective masses $(m_0,m)$; the optimal operation point is denoted by a white $\times$ marker. We notice that the system is much more sensitive to changes of the effective mass $m_0$ of the background than to that of the shell medium $m$; such a property is related to the fact that the effective mass $m_0$ of the host is usually much larger than the corresponding one of the cylindrical layer. Importantly, a cladding with larger effective mass $m$ require a slightly smaller $m_0$ of the background to serve with the same success the sinking operation. 

In Fig. \ref{fig:Fig4b}, we represent the performance $Q$ across the corresponding map of potential energies and realize that the design is quite robust to random changes since the regarded range of the background parameter $V'_0/V_0$ is hundred times larger compared to the respective $m'_0/m_0$ one of Fig. \ref{fig:Fig4a} while the focusing deterioration is much milder. It is again noted that the examined nanotube continues to work efficiently as a sink for a broad range of potentials $V$ values but is more susceptible to misengineered background values $V_0$, as happens with the effective masses in Fig. \ref{fig:Fig4a}. Interestingly, the global maximum does not coincide with the one dictated by the optimization process since, in our scheme, the potentials $(V_0,V)$ are not taken as free continuous variables but possess discrete values provided via experimental measurements for each medium; similar considerations hold for Fig. \ref{fig:Fig4a}.     

\begin{figure}[ht!]
\centering
\subfigure[]{\includegraphics[width=5.5cm]{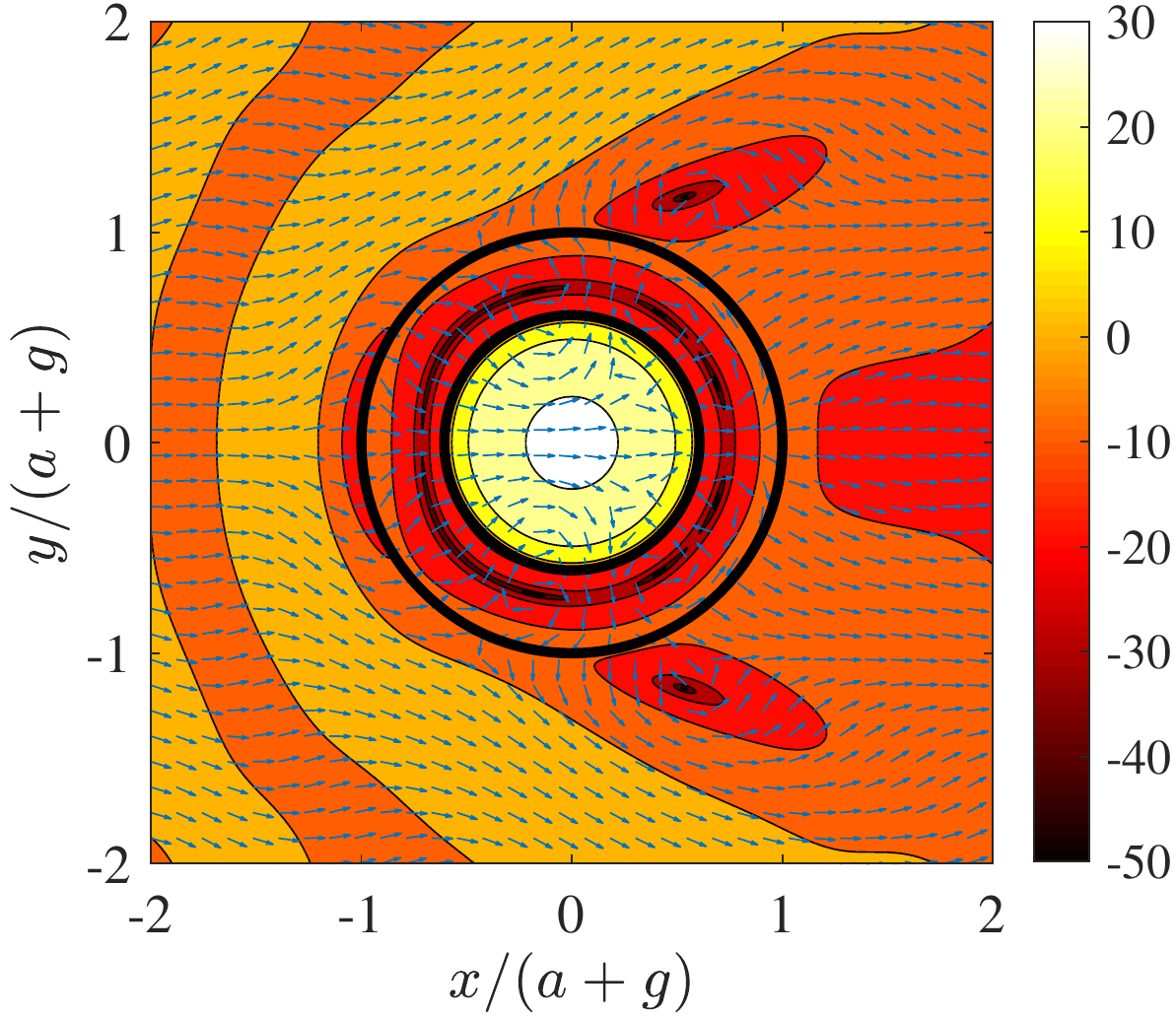}
   \label{fig:Fig5a}}
\subfigure[]{\includegraphics[width=5.5cm]{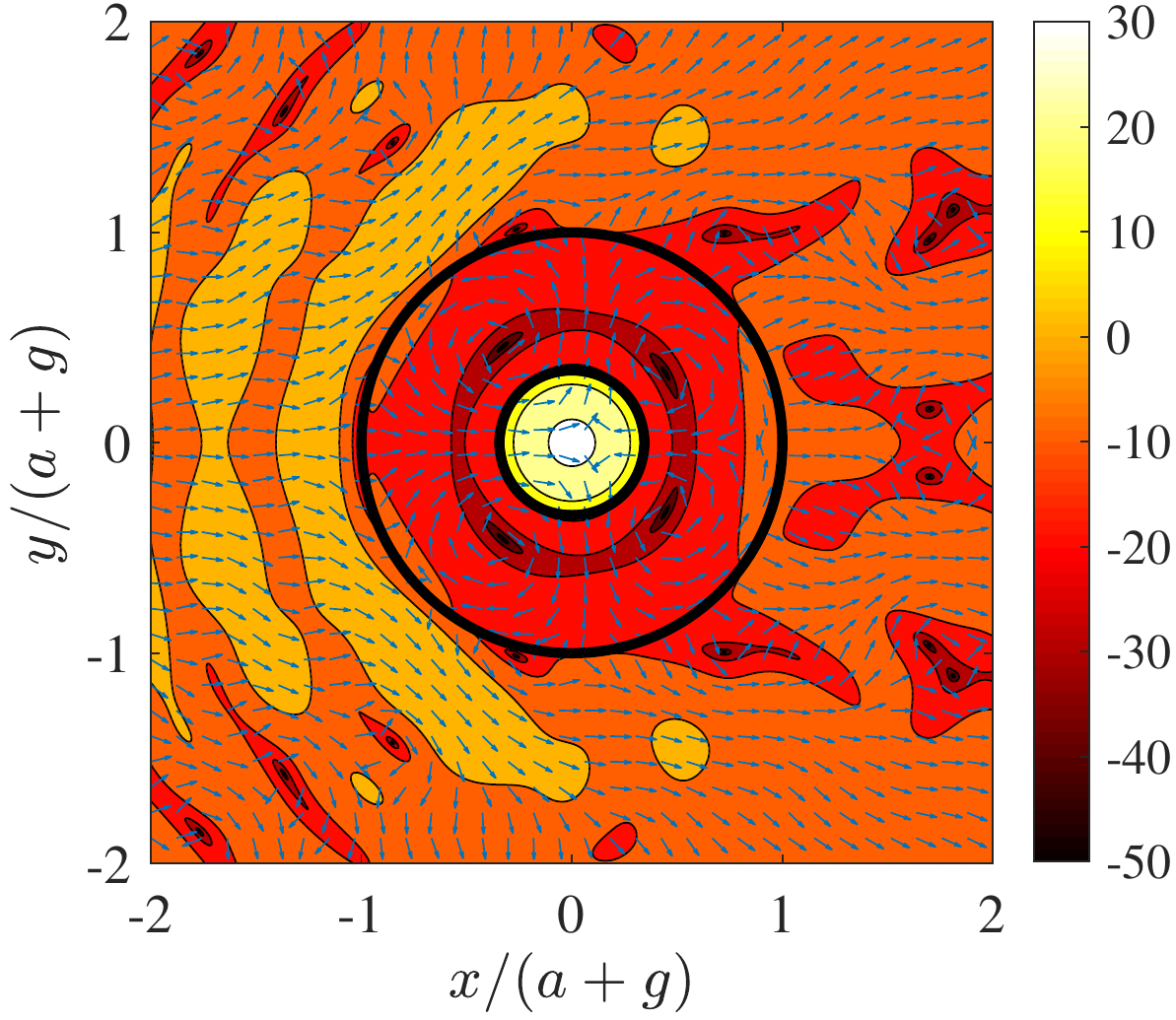}
   \label{fig:Fig5b}}
\caption{Spatial distribution of the (unnormalized) probability $|\Psi(x,y)|^2$ in dB in the vicinity of the considered quantum particle sink. (a) InSb shell in SiC, low energy: $E=0.1$ eV, (b) InSb shell in diamond, high energy: $E=1$ eV. The blue arrows correspond to the probability current $\textbf{J}$ from \eqref{ProbabilityCurrent} and the black circles denote the nanotubes boundaries.}
\label{fig:Figs5}
\end{figure}

\subsection{Focused Quantum Signals}
In Fig. \ref{fig:Figs5}, to demonstrate the efficiency of the proposed designs, we show the spatial variation of the probability $|\Psi(x,y)|^2$ of finding a quantum particle at a specific point $(x,y)$. On top of the maps, we depict the vector field of the probability current defined by:
\begin{eqnarray}
\textbf{J}(\textbf{r})\!=\!-\frac{\hbar}{m(\textbf{r})}\Im\left[ \Psi^*(\textbf{r}) \nabla\Psi(\textbf{r}) \right],
\label{ProbabilityCurrent}
\end{eqnarray}
which is the closest quantity we have to describe the velocity of quantum particles.

In Fig. \ref{fig:Fig5a}, we consider an optimal design working for low-energy particles (InSb shell in SiC host from Fig. \ref{fig:Fig2a}) and the unnormalized probability $|\Psi(x,y)|^2$ internally to the nanotube is around 1000 times larger than the (unitary) probability of the impinging wave in the unbounded homogeneous background medium. We notice that the represented quantity drops dramatically into the shell while multiple vortices of the probability current are observed around the points with $\Psi(x,y)\rightarrow 0$. Indeed, the electrons will inevitably choose an outgoing spiral direction from the eyes of the whirlpools since they cannot exist at a point with zero probability.

In Fig. \ref{fig:Fig5b}, we examine another optimized configuration but, this time, operated for high-energy electrons (InSb shell in diamond host from Fig. \ref{fig:Fig2b}). It is clear that the perturbation around the structure is more significant and the developed dynamics richer than in the case of Fig. \ref{fig:Fig5a}. In addition, the regions of vanishing wave function in the vicinity of the core, force an inversion of the probability current which, in turn, creates a larger particle concentration (up to 30 dB enhancement) into the sink.

\section{Conclusions}
Thin cylindrical layers are found suitable to work as sinks for the impinging quantum particles by concentrating the probability of finding them, into their cores. Every single combination of quantum materials picked from a long list is tried and the considered structures are carefully optimized; as a result, several ultra-efficient setups are reported to achieve enhancement of the trapping effect by two to three orders of magnitude. The proposed designs are found to be very selective with respect to the energy of the incoming beams but quite robust to changes in the effective parameters describing the background host. The spatial distributions of the wave functions and the probability currents demonstrate the focusing of matter waves and unveil the characteristics of the sustained resonances.

An interesting expansion of the present work would be to consider multiple nanotubes working collectively to create the desired scattering beam patterns into the background medium they are embedded \cite{SPIEConf}. The employed materials can also exhibit nonlinearities \cite{MyJ121} or possess gain \cite{MyJ117} that lead to much more complex dynamic schemes calling for more cumbersome mathematical manipulation but also allowing for more sophisticated utilities. All these enriched versions have been solved for electromagnetic waves and thus can be readily applied in the case of matter waves by serving the translation of photonic concepts into the quantum arena. 

\section*{Acknowledgement}    
This work was partially supported by Nazarbayev University Faculty Development Competitive Research Grant No. 021220FD4051 (\textit{``Optimal design of photonic and quantum metamaterials''}).

\bibliographystyle{plain}

\end{document}